\documentclass[aps,pra,twocolumn,showpacs,superscriptaddress,groupedaddress]{revtex4-2}  % for review and submission

\usepackage{graphicx}  % needed for figures
\usepackage{dcolumn}   % needed for some tables
\usepackage[colorlinks = true,
            linkcolor = blue,
            urlcolor  = blue,
            citecolor = blue,
            anchorcolor = blue]{hyperref}
\begin{document}
	
	% The following information is for internal review, please remove them for submission
	%\widetext
	%\leftline{Version xx as of \today}
	%\leftline{Primary authors: Joe E. Physics}
	%\leftline{To be submitted to (PRL, PRD-RC, PRD, PLB; choose one.)}
	%\leftline{Comment to {\tt d0-run2eb-nnn@fnal.gov} by xxx, yyy}
	%\centerline{\em D\O\ INTERNAL DOCUMENT -- NOT FOR PUBLIC DISTRIBUTION}
	
	% the following line is for submission, including submission to the arXiv!!
	%\hspace{5.2in} \mbox{Fermilab-Pub-04/xxx-E}
	
\title{Frequency-comb-induced radiation pressure force in dense atomic clouds}

\author{M.~Kruljac}
\affiliation{Institute of Physics, Bijeni\v{c}ka cesta 46, 10000 Zagreb, Croatia}	
\author{D.~Buhin}
\affiliation{Institute of Physics, Bijeni\v{c}ka cesta 46, 10000 Zagreb, Croatia}
\author{D.~Kova\v{c}i\'{c}}
\affiliation{Institute of Physics, Bijeni\v{c}ka cesta 46, 10000 Zagreb, Croatia}
\author{V.~Vuli\'{c}}
\affiliation{Institute of Physics, Bijeni\v{c}ka cesta 46, 10000 Zagreb, Croatia}
\author{D.~Aumiler}
\affiliation{Institute of Physics, Bijeni\v{c}ka cesta 46, 10000 Zagreb, Croatia}
\author{T.~Ban}
\affiliation{Institute of Physics, Bijeni\v{c}ka cesta 46, 10000 Zagreb, Croatia}
\email{ticijana@ifs.hr}

\date{\today}

\begin{abstract}
	We investigate the frequency comb induced radiation pressure force acting on a cloud of cold $^{87}$Rb atoms.
	Reduction and spectral broadening of the frequency comb force are observed as the cloud's optical thickness is increased.
	Since the radiation pressure force is uniquely determined by light scattered by an atomic cloud, we discuss different scattering mechanisms, and point to the shadow effect as the dominant mechanism affecting FC-induced force in resonantly excited dense atomic clouds.
	Our results improve the understanding of the interaction of frequency comb light with many-atom ensembles, which is essential for novel frequency comb applications in simultaneous multi-species cooling, multi-mode quantum memories, and multi-mode atom-light interfaces. 
\end{abstract}

\maketitle

\section{\label{sec:level1}Introduction}

Optical frequency combs (FCs) have become an essential source of light in applications ranging from metrology \cite{ye2003, rosenband2008} and high-resolution spectroscopy \cite{diddams2007, maslowski2014, picque2019} to precision ranging \cite{minoshima2000} and calibration of atomic spectrographs \cite{murphy2007}.
In recent years, the applications of FCs have expanded to laser cooling and quantum communication.
In quantum communication, the FC offers tremendous potential for the realization of multi-mode nonclassical light \cite{cai2017, kues2017, reimer2016, maltese2020} and multi-mode quantum memories \cite{kresic2019, main2020}. 
Regarding laser cooling applications, FC cooling of ions \cite{udem2016, ip2018}, neutral atoms \cite{cambell2016, santic2019}, and simultaneous dual-species FC cooling \cite{buhin2020} have recently been demonstrated.
In addition, a recent theoretical proposal envisions enhanced cavity cooling and complex self-ordering patterns when an optical resonator filled with a cold atomic gas is pumped by a multitude of FC comb modes \cite{torggler2020}.    
For these novel FC applications, that are at the core of emerging quantum technologies, it is necessary to fully understand the interaction of FC light with many-atom ensembles, and in particular the FC-induced radiation pressure force that is induced on an atomic ensemble by the FC excitation.

Investigation of light scattered from an ensemble of cold atoms illuminated by a continuous wave (cw) laser has been an extremely fruitful platform for studying light-matter interactions \cite{bromley2016, pellegrino2014, jenkins2016, balik2013, bloch2020}.
Cooperative scattering by an ensemble of resonant systems has been studied in detail by R. Dicke \cite{dicke1954} and has led to understanding of superradiance and collective level shifts. 
Several experiments studied the radiation pressure force exerted by a cw laser on a cold atomic cloud in order to capture the signature of cooperative effects \cite{bienaime2010, bux2010, courteille2010, chabe2014}, since the force is uniquely determined by the light scattered from the atomic cloud. 
Bienaime et al. used a timed-Dicke state (TDS) approach to calculate the cooperative radiation pressure force acting on a cloud of cold $^{87}$Rb atoms \cite{bienaime2010}. 
Excellent agreement of the experiment and calculations indicated that the radiation pressure force could be used as a new tool for the observation of cooperativity.
That was a very attractive idea which led to a series of new studies. 
The later studies, however, indicated that specific effects observed in radiation pressure force may not always be a signature of cooperativity, but a result of different incoherent scattering mechanisms such as attenuation of the probe light, diffraction and refraction, multiple scattering, etc. \cite{bachelard2016, guerin2016}. 

In this paper we investigate the FC-induced radiation pressure force acting on a cloud of cold $^{87}$Rb atoms released from a magneto-optical trap (MOT).
Reduction and spectral broadening of the frequency comb force are observed as the cloud's optical thickness is increased.
Based on the theoretical models developed for the cw radiation pressure force \cite{bachelard2016}, we discuss the role of diffuse, Mie, and cooperative scattering in the observed FC-induced force.
We conclude that the FC-induced force is predominately affected by the progressive attenuation of the light intensity within the cloud due to diffuse scattering of light, i.e. we identify the so-called shadow effect as the dominant mechanism affecting the FC-induced force in dense atomic clouds.    
Our results support the considerations in \cite{bachelard2016} for the case of cw-induced force and small optical thicknesses, thus verifying the analogy between the interaction of a FC light and a cw laser light with atomic ensembles.

\section{Experiment}
\begin{figure}[ht!]
	\centerline{
		\mbox{\includegraphics[width=0.45\textwidth]{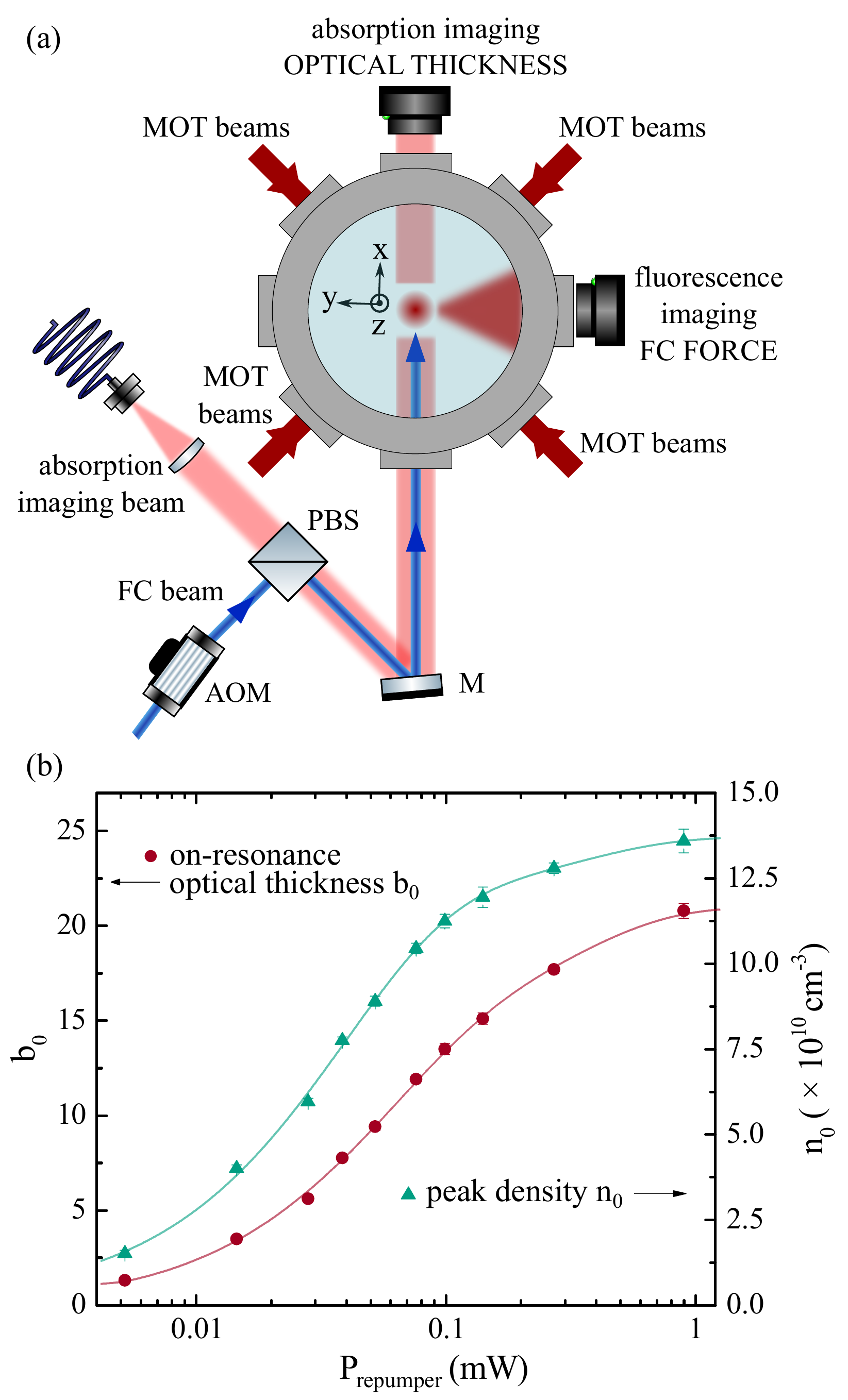}}
	}
	\caption{(a) A simplified experimental scheme. Two pairs of MOT beams are shown, while the third pair is propagating along the $z$ axis. The absorption imaging beam and the FC beam are co-propagated in the $x$-axis. The optical thickness is measured using the absorption imaging camera, while the FC force is measured using the fluorescence imaging camera. During the measurement of optical thickness, the FC beam is blocked using an AOM. M is a mirror and PBS is a polarizing beam splitter. (b) On-resonance optical thickness (red circles) and cloud peak density (green triangles) as a function of the repumper laser power during the MOT loading stage. Solid lines represent a guide to the eye.
	}
	\label{Fig1}
\end{figure}

A simplified scheme of the experimental setup for the preparation of a cold $^{87}$Rb cloud and its characterization using absorption imaging, as well as the setup for FC force measurement using fluorescence imaging is shown in Fig. \ref{Fig1}(a).

%%%%%%%%%%%%%%%%%%%%%%%%%%%%%%%%%%%%%%%%%%%%%%%%%%%%%%%
%%%%%%%%%%%%%%%%%% FIGURE 1 %%%%%%%%%%%%%%%%%%%%%%%%%%%%%%%%%%%%%%%%%%%%%%%%%%%%%%%%%%%%%%%%%%%%%%%%%%%%%%%%%%

%%%%%%%%%%%%%%%%%%%%%%%%%%%%%%%%%%%%%%%%%%%%%%%%%%%%%%%
%%%%%%%%%%%%%%%%%%%%%%%%%%%%%%%%%%%%%%%%%%%%%%%%%%%%%%%

\textbf{Preparation of a cloud of cold atoms}.
A cold $^{87}$Rb cloud is loaded from a background vapor in a stainless steel chamber using a standard six-beam configuration. 
The preparation of a cold cloud of a given optical thickness is achieved in three consecutive stages: MOT loading, temporal dark MOT, and repumping stage.
In the first stage, we load the MOT for 6 s, with the cooling laser detuned $-3.5\Gamma$ from the $^{87}$Rb $|5S_{1/2};F=2\rangle \rightarrow |5P_{3/2};F'=3\rangle$ transition, and the repumper laser in resonance with the $|5S_{1/2};F=1\rangle \rightarrow |5P_{3/2};F'=2\rangle$ transition, generating a cloud of $\approx$ $4 \cdot 10^7$ atoms at a temperature of around 50 $\mu$K, and a $1/e^2$ radius of $\approx$ 0.8 mm. 
Here $\Gamma=2\pi\cdot$ 6.07 MHz is the  natural linewidth of the $|5S_{1/2}\rangle \rightarrow |5P_{3/2}\rangle$ transition \cite{steck1}.
In the second stage, we apply a 15 ms long temporal dark MOT, where we reduce the power of the repumper laser to 10 $\mu$W and the detuning of the cooling laser to $-2\Gamma$, leaving other parameters unchanged.
As a result, the atoms are pumped into the $F=1$ ground level, which causes an increase of the cloud density and, consequently, of the optical thickness. 
Finally, we increase the power of the repumper laser to 1.5 mW and tune the cooling laser to $-7\Gamma$, in order to re-cool and compress the cloud into a spherical shape while pumping all the atoms back into the $F=2$ ground state in the third repumping stage that lasts 1 ms. 

After preparing the cold cloud, we measure its optical thickness, $b(y,z)$, using the standard absorption imaging technique on the $|5S_{1/2};F=2\rangle \rightarrow |5P_{3/2};F'=3\rangle$ transition.
For an atomic cloud of a Gaussian density distribution, $b(y,z)$ will also have a Gaussian shape.
By fitting a 2D Gaussian to the measured $b(y,z)$, we extract the optical thickness at the centre of the cloud, $b_{peak}$.
On-resonance optical thickness, $b_0$, is calculated using $b_0 = b_{peak} \cdot \left(1+4 \delta_{img}^2/\Gamma^2\right)$, where $\delta_{img}$ is the detuning of the probe laser frequency used for absorption imaging.
On-resonance optical thickness is defined as $b_0=\sigma_0\int_{-\infty}^{\infty} n(x,y$=$z$=$0)\,dx$, where $\sigma_0$ is on-resonance cross section \cite{steck1}, and $n(x,y,z)$ is the spatial density of the cloud.

In order to vary the optical thickness of the cloud, we change the power of the repumper laser in the MOT loading stage, leaving the dark MOT and the repumping stage parameters unchanged. 
This change in the loading stage also affects other cloud parameters, such as size, number of atoms and temperature.
This does not affect the accuracy of optical thickness determination since it is measured directly by absorption imaging; nevertheless, a detailed characterization of all cloud parameters has been made.
In Fig. \ref{Fig1}(b), the peak density, $n_0=n(x$=$y$=$z$=$0)$, and on-resonance optical thickness, $b_0$, are shown as a function of the repumper laser power in the MOT loading stage. 
For the given range of powers, the cloud temperature varies from 35 $\mu$K to 75 $\mu$K, measured using a standard time-of-flight (TOF) technique.

\textbf{FC force measurement}.
The FC is generated by frequency doubling an Er:fiber mode-locked femtosecond laser (TOPTICA FFS) operating at 1560 nm with a repetition rate of $f_{rep}$=80.495 MHz. 
The frequency-doubled spectrum is centered around 780 nm with a FWHM of about 5 nm and a total output power of 76 mW. 
The FC spectrum consists of a series of sharp lines, i.e. comb modes \cite{cundiff2005}. 
The optical frequency of the $n$-th comb mode is given by $f_n=n \cdot f_{\text{rep}}+f_0$, where $f_0$ is the offset frequency.
In our experiment, we actively stabilize $f_{\text{rep}}$ and $f_n$ by giving feedback to the cavity length and pump power of the mode-locked laser, thus indirectly fixing $f_0$. 
The frequency of the $n$-th comb mode is varied by scanning $f_0$ while keeping $f_{\text{rep}}$ fixed. 
A detailed description of the FC stabilization and scanning scheme is presented in our recent papers \cite{santic2019, buhin2020}.

The measurement sequence starts after the preparation of a cloud of a given optical thickness, and is similar to the one described in our recent works \cite{santic2019, buhin2020}.
At $t=0$ we turn off the MOT cooling beams and switch on the linearly polarized FC beam. The total power of the FC beam on the atoms is 25 mW and the beam size ($1/e^2$) is 4.5 mm, resulting in the power and intensity per comb mode of about 0.75 $\mu$W and 9 $\mu$W/cm$^{2}$, respectively. The MOT repumper laser is left on to continuously pump the atoms out of the $|5S_{1/2};F=1\rangle$ ground level and has no measurable mechanical effect. The quadrupole magnetic field is also left on. We let the comb interact with the cold cloud for 0.5 ms.
During this time the center of mass (CM) of the cloud accelerates in the FC beam direction ($+x$-direction) due to the FC force.
The FC and repumper beams are then switched off, and the cloud expands freely for a variable time, after which we switch on the MOT cooling beams for 0.15 ms and image the cloud's fluorescence with a camera to determine its CM displacement. 

It is worth noting here that the approaches to change the optical thickness of the cloud by changing the repumper laser power immediately after the dark MOT stage used in \cite{bienaime2010}, and by changing the cloud's expansion time before interaction as used in \cite{bromley2016}, are not applicable in our case of the FC excitation.  
In the first approach, only a fraction of atoms are transferred from $|5S_{1/2};F=1\rangle$ to $|5S_{1/2};F=2\rangle$ ground level after the dark MOT, depending on the repumper laser power. 
Atoms remaining in the $|5S_{1/2};F=1\rangle$ level and atoms in $|5S_{1/2};F=2\rangle$ could be simultaneously excited by different comb modes, which would result in a complex lineshape of the measured FC force.
In the second approach, the size of the FC beam should be bigger than the size of the expanding cloud, which cannot be achieved in our setup due to the low power per comb mode.

\section{Results and discussion}
\subsection{FC force as a function of cloud density}
\begin{figure}[ht!]
	\centerline{
		\mbox{\includegraphics[width=0.47\textwidth]{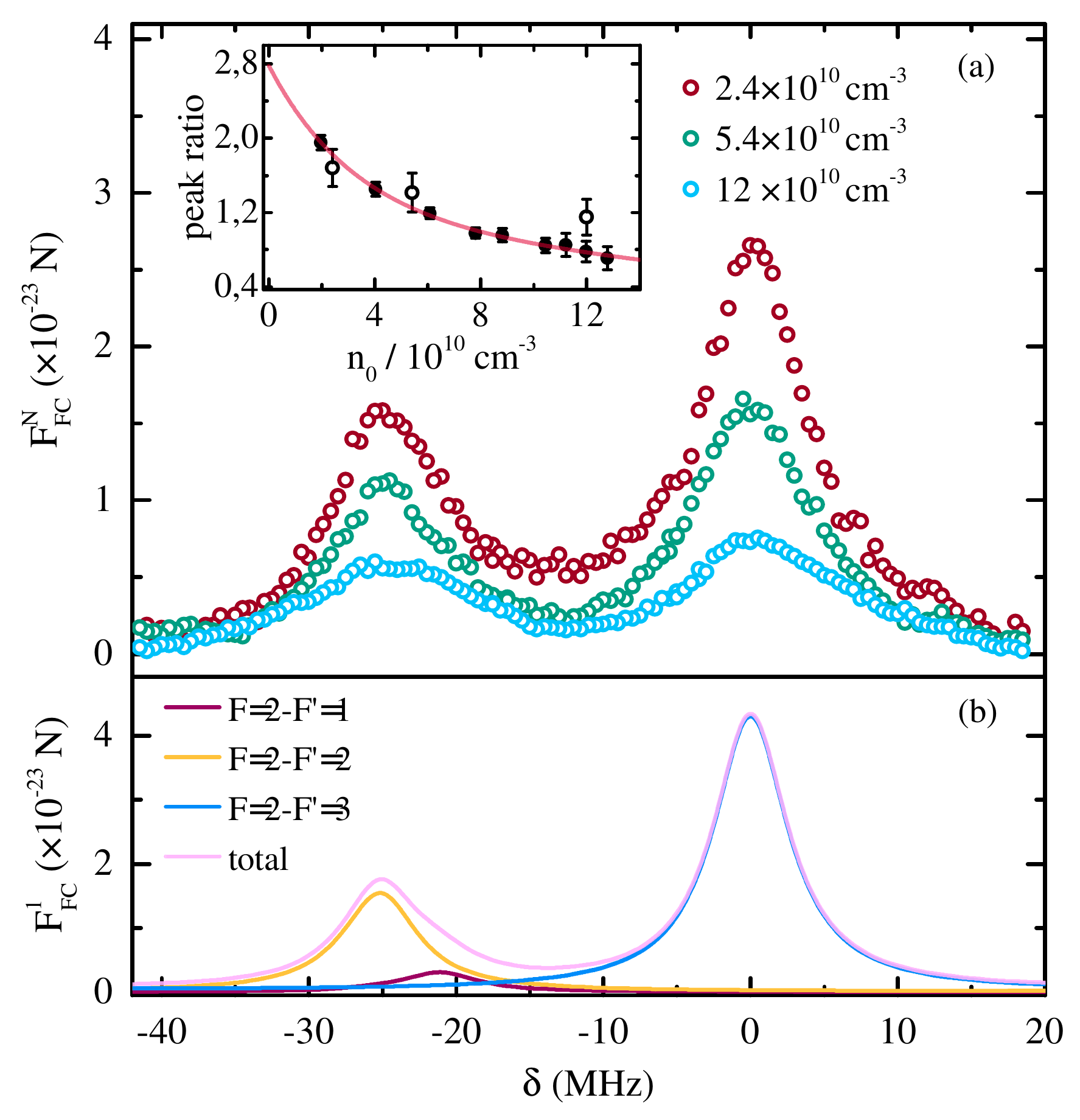}}
	}
	\caption{(a) Measured FC force,  $F^N_{FC}$, as a function of the FC detuning $\delta$, for different peak densities $n_0$. Inset shows the ratio of the FC peak forces at $\delta=0$ and $\delta\approx -25.5$ MHz where the symbols are experimental data and the line represents a guide to the eye. Full circles correspond to averaged multiple scans, as described in the experimental section. Empty circles correspond to peak ratios of the scans shown in (a), which were taken without averaging and thus have larger errors. (b) Calculated FC force,  $F^1_{FC}$, as a function of the FC detuning $\delta$. The total FC force (violet line) is obtained by summing the force contributions from three $|5S_{1/2};F=2\rangle \rightarrow |5P_{3/2};F'=1,2,3\rangle$ hyperfine transitions \cite{santic2019}. 
	}
	\label{Fig2}
\end{figure}

In Fig. \ref{Fig2}(a) we show the measured FC force, $F^N_{FC}$, as a function of the FC detuning $\delta$, which we define as the detuning of the $n$-th comb mode from the $|5S_{1/2};F=2\rangle \rightarrow |5P_{3/2};F'=3\rangle$ transition, for different peak cloud densities, $n_0$. Due to the nature of the comb spectrum, the FC radiation pressure force is periodic with respect to the comb detuning with a period equal to $f_{\text{rep}}$. 
Two distinct peaks appear in one $f_{\text{rep}}$ scan, reflecting the interaction with three comb modes, as explained in detail in our recent work \cite{santic2019}.
The peak at $\delta=0$ is due to the $n$-th comb mode being in resonance with the $|5S_{1/2};F=2\rangle \rightarrow |5P_{3/2};F'=3\rangle$ transition, whereas the peak at $\delta\approx -25.5$ MHz is due to the $(n-3)$-rd mode being in resonance with the $|5S_{1/2};F=2\rangle \rightarrow |5P_{3/2};F'=2\rangle$ transition and the $(n-5)$-th mode with the $|5S_{1/2};F=2\rangle \rightarrow |5P_{3/2};F'=1\rangle$ transition.
For completeness, in Fig. \ref{Fig2}(b) we show the calculated FC force, $F^1_{FC}$, obtained by summing the contributions from three hyperfine transitions. 
The FC force is calculated for a single atom, and the details of the calculation can be found in \cite{santic2019}.   

As the cloud density increases, broadening and reduction of both FC force peaks are observed.
In addition, the ratio of the peaks at $\delta=0$ and $\delta\approx -25.5$ MHz decreases with increasing density, as can be seen from the inset in Fig. \ref{Fig2}(a).
The peak ratio of 2.8 is expected when $n_0$ approaches zero, as it reflects the ratio of the $|5S_{1/2};F=2\rangle \rightarrow |5P_{3/2};F'=3\rangle$ and  $|5S_{1/2};F=2\rangle \rightarrow |5P_{3/2};F'=2\rangle$ transition dipole moments \cite{steck1}. 
This can be easily understood given the well-known result that the force broadening and reduction due to collective effects in many-atom ensembles depend on the optical thickness rather than the density \cite{zhu2016}. 
Since the optical thickness is defined through the cross section $\sigma_0 = h \omega \Gamma / (2I_{sat})$, where $I_{sat}$ is the saturation intensity that depends on the dipole moment of the relevant transition \cite{steck1}, the two peaks have different optical thicknesses for a given density, and therefore different factors of force reduction, which directly affects the peak ratio.
In the following sections we will therefore present and analyze the dependence of the FC force on the optical thickness for each force peak separately.

%%%%%%%%%%%%%%%%%%%%%%%%% FIGURE 2 %%%%%%%%%%%%%%%%%%%%%%%%%%%%%%
%%%%%%%%%%%%%%%%%%%%%%%%%%%%%%%%%%%%%%%%%%%%%%%%%%%%%%%

\subsection{FC force as a function of cloud optical thickness}

\begin{figure}[ht!]
	\centerline{
		\mbox{\includegraphics[width=0.46\textwidth]{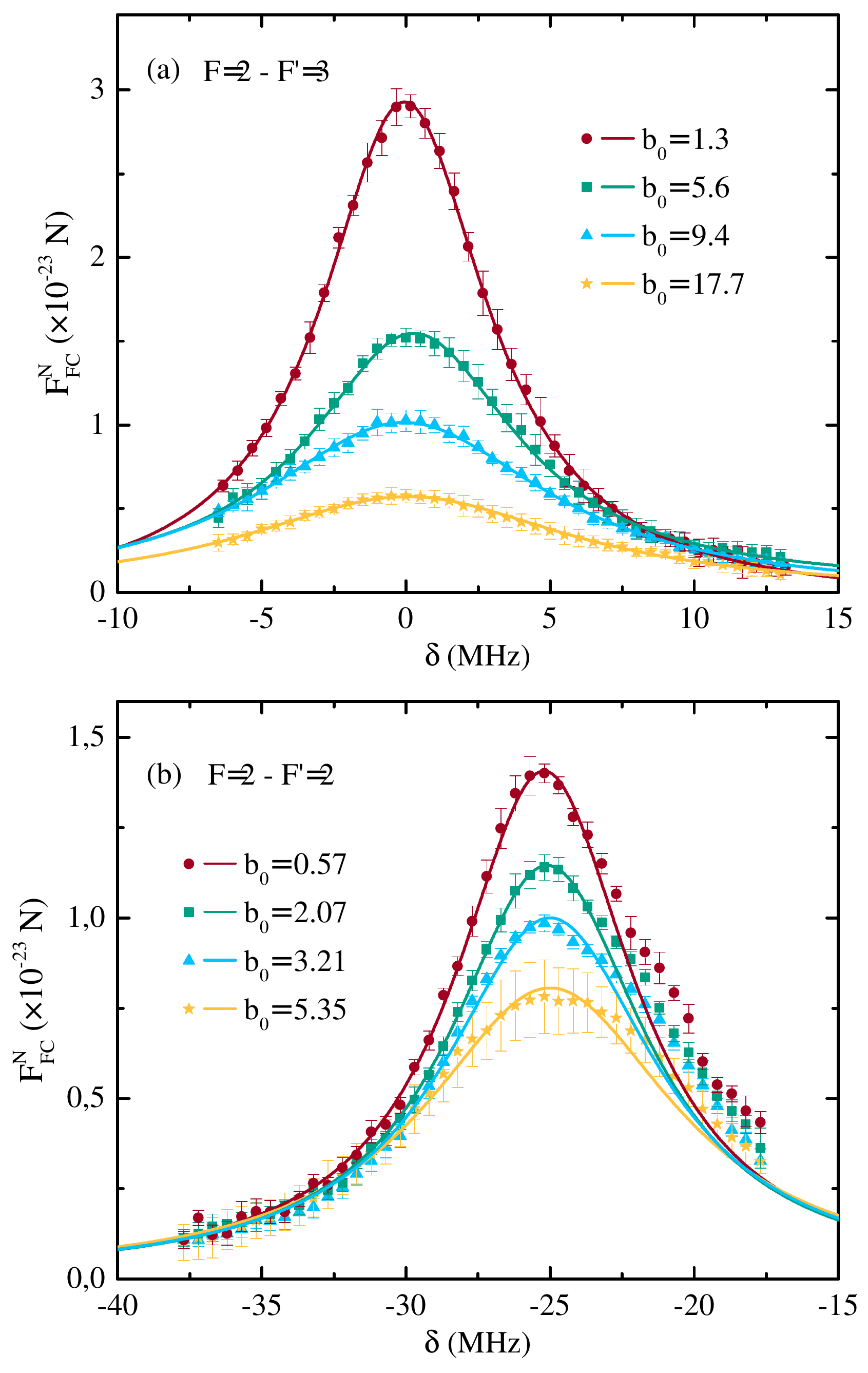}}
	}
	\caption{Measured FC force (symbols) as a function of detuning $\delta$ for different optical thicknesses $b_0$. (a) FC force is due to the $n$-th comb mode being in resonance with the $|5S_{1/2};F=2\rangle \rightarrow |5P_{3/2};F'=3\rangle$ transition. (b) FC force is due to the $(n-3)$-rd mode being in resonance with the $|5S_{1/2};F=2\rangle \rightarrow |5P_{3/2};F'=2\rangle$ transition and the $(n-5)$-th mode with the $|5S_{1/2};F=2\rangle \rightarrow |5P_{3/2};F'=1\rangle$ transition. A solid line shows a Lorentzian fit to the experimental data.  
	}
	\label{Fig3}
\end{figure}

In Fig. \ref{Fig3} we show the measured FC force,  $F^N_{FC}$, as a function of the FC detuning $\delta$ for different on-resonance optical thicknesses $b_0$.
In the case of the $|5S_{1/2};F=2\rangle \rightarrow |5P_{3/2};F'=3\rangle$ transition, $b_0$ is measured directly as described in Sec. II, and is divided by 2.8 to obtain $b_0$ relevant for the $|5S_{1/2};F=2\rangle \rightarrow |5P_{3/2};F'=2\rangle$ transition.

%%%%%%%%%%%%%%%%%% FIGURE 3 %%%%%%%%%%%%%%%%%%%%%%%%%%%%%%%%%%%%%%%%%%%%%%%%%%%%%
%%%%%%%%%%%%%%%%%%%%%%%%%%%%%%%%%%%%%%%%%%%%%%%%%%%%%%%%%%%%%%%%%%%%%%%%%%%%%%%%%
%%%%%%%%%%%%%%%%%%%%%%%%%%%%%%%%%%%%%%%%%%%%%%%%%%%%%%%%%%%%%%%%%%%%%%%%%%%%%%%%% 

The measured FC forces, arising from the $|5S_{1/2};F=2\rangle \rightarrow |5P_{3/2};F'=3\rangle$ transition show a Lorentzian line shape in the whole range of measured $b_0$, Fig. \ref{Fig3}(a). % for $1.3 \leq b_0 \leq 17.7$, Fig. \ref{Fig3}(a).
In the case of the $|5S_{1/2};F=2\rangle \rightarrow |5P_{3/2};F'=2\rangle$ transition, the FC forces deviate from the Lorentzian line shape, Fig. \ref{Fig3}(b), due to the $|5S_{1/2};F=2\rangle \rightarrow |5P_{3/2};F'=1\rangle$ FC force contribution positioned in the blue wing of the peak, as indicated in Fig. \ref{Fig2}(b).

For a given $b_0$, a Lorentzian function is fitted to the experimental data.
For the $|5S_{1/2};F=2\rangle \rightarrow |5P_{3/2};F'=2\rangle$ transition, we fit only to the data on the red side of the curve, where the influence of the $|5S_{1/2};F=2\rangle \rightarrow |5P_{3/2};F'=1\rangle$ transition is negligible.
While the FC force offset should be zero, experimentally we see a small offset due to inaccuracies in determination of the initial and final position of the cloud's CM, from which the force is determined.
The small FC force offset is subtracted from all experimental data shown in the Figs. \ref{Fig3} and \ref{Fig4}.

The FC force broadening and reduction are clearly observed for both peaks shown in Figs. \ref{Fig3}(a) and \ref{Fig3}(b) and are presented in more details in Fig. \ref{Fig4}.

\subsection{FC force broadening and reduction}
\begin{figure}[ht!]
	\centerline{
		\mbox{\includegraphics[width=0.46\textwidth]{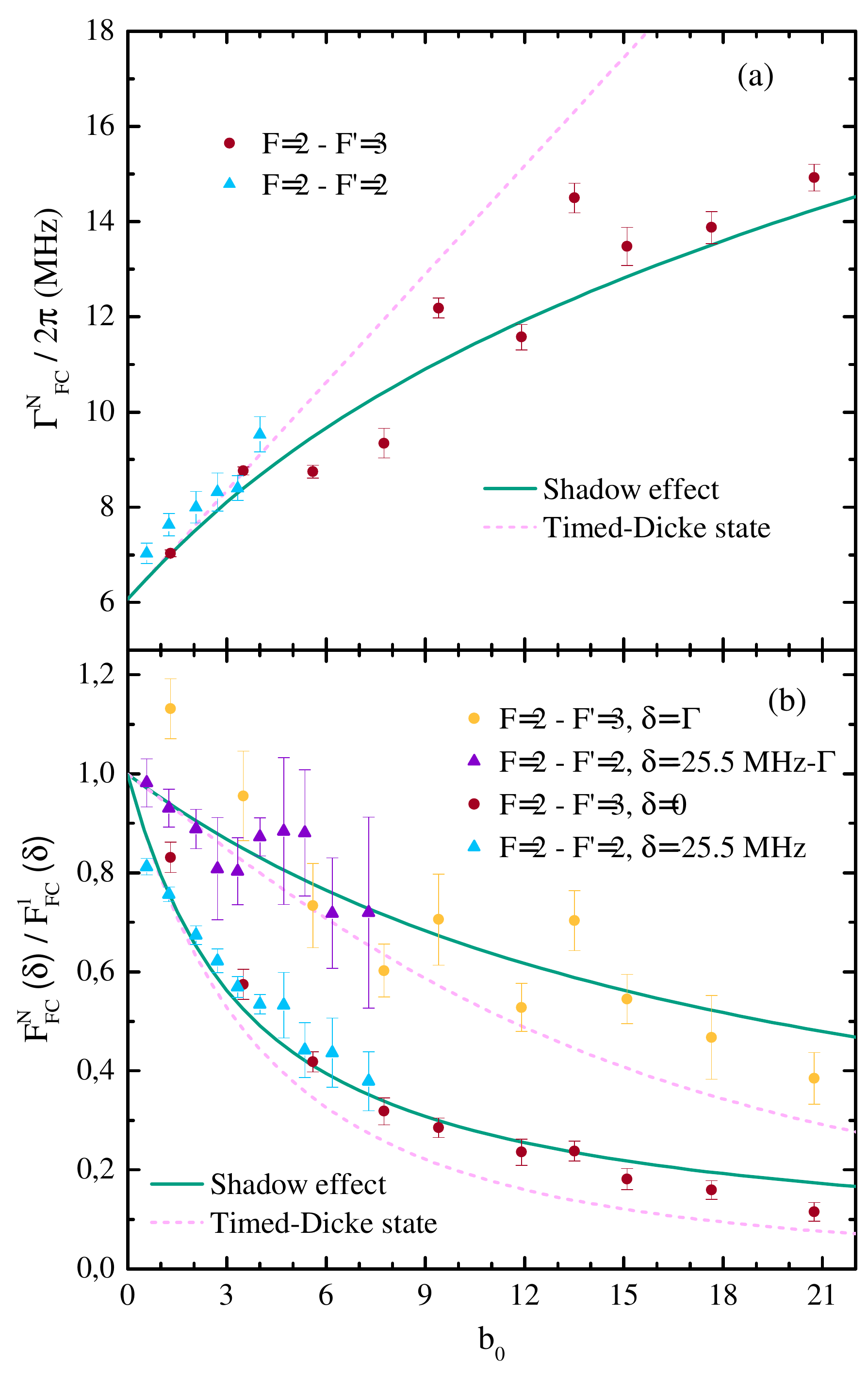}}
	}
	\caption{(a) Measured (symbols) and calculated FC force broadening in the presence of shadow (solid line) and cooperative (dashed line) effects, using Eq. (\ref{force_shadow}) and (\ref{force_cooperative}), respectively, as a function of $b_0$. (b) Measured (symbols) and calculated FC force reduction as a function of $b_0$ at $\delta=0$ and $\delta= - \Gamma$ in the case of $|5S_{1/2};F=2\rangle \rightarrow |5P_{3/2};F'=3\rangle$ transition, and $\delta=-25.5$ MHz and $\delta=-25.5$ MHz$-\Gamma$ in the case of $|5S_{1/2};F=2\rangle \rightarrow |5P_{3/2};F'=2\rangle$ transition.
	}
	\label{Fig4}
\end{figure}

In Fig. \ref{Fig4}(a) we show the measured (symbols) FC force linewidths, $\Gamma_{FC}^N$, as a function of $b_0$. 
For a given $b_0$, $\Gamma_{FC}^N$ is obtained from the fit of a Lorentzian function to the measured FC force spectra as shown in Fig. \ref{Fig3}(a) for $|5S_{1/2};F=2\rangle \rightarrow |5P_{3/2};F'=3\rangle$ transition and in Fig. \ref{Fig3}(b) for $|5S_{1/2};F=2\rangle \rightarrow |5P_{3/2};F'=2\rangle$ transitions.
We observe an increase of the FC force linewidth with increasing $b_0$.
For small $b_0$ the increase is linear, while the curve starts to flatten as the $b_0$ is increased. 
In the limit $b_0\rightarrow 0$, the FC linewidth of $\Gamma=2\pi\cdot$ 6.07 MHz is expected, as it reflects the natural linewidth of the $^{87}$Rb $|5S_{1/2}\rangle \rightarrow |5P_{3/2}\rangle$ transition \cite{steck1}.
For the largest $b_0=20.8$ achieved in the experiment, the FC force linewidth of $2.5\:\Gamma$ is measured.

In Fig. \ref{Fig4}(b) we show the measured (symbols) reduction of the FC force, $F^N_{FC}(\delta)/F^1_{FC}(\delta)$, as a function of $b_0$.
$F^N_{FC}(\delta)$ are obtained from the measured FC force spectra as shown in Fig. \ref{Fig3}(a) (for $\delta=0$ and $\delta= - \Gamma$) in the case of $|5S_{1/2};F=2\rangle \rightarrow |5P_{3/2};F'=3\rangle$ transition, and in Fig. \ref{Fig3}(b) (for $\delta=-25.5$ MHz and $\delta=-25.5$ MHz$-\Gamma$) in the case of $|5S_{1/2};F=2\rangle \rightarrow |5P_{3/2};F'=2\rangle$ transition.
The single atom force $F^1_{FC}(\delta)$ is obtained by fitting Eq. (\ref{force_shadow}) to the measured data with $F^1_{FC}(\delta)$ as a free parameter (see the following paragraph for details). 
A reduction of the FC force with increasing $b_0$ is observed. 
The force reduction is larger when the relevant comb mode is resonant with a given atomic transition, i.e. when the $n$-th comb mode in resonance with the $|5S_{1/2};F=2\rangle \rightarrow |5P_{3/2};F'=3\rangle$ transition ($\delta=0$), and the $(n-3)$-rd with the $|5S_{1/2};F=2\rangle \rightarrow |5P_{3/2};F'=2\rangle$ transition ($\delta=-25.5$ MHz).
For the largest $b_0=20.8$ achieved in the experiment, the FC force reduction of almost $90\%$ is measured. 

%%%%%%%%%%%%%%%%%% FIGURE 4 %%%%%%%%%%%%%%%%%%%%%%%%%%%%%%%%%%%%%%%%%%%%%%%%%%%%%
%%%%%%%%%%%%%%%%%%%%%%%%%%%%%%%%%%%%%%%%%%%%%%%%%%%%%%%%%%%%%%%%%%%%%%%%%%%%%%%%%
%%%%%%%%%%%%%%%%%%%%%%%%%%%%%%%%%%%%%%%%%%%%%%%%%%%%%%%%%%%%%%%%%%%%%%%%%%%%%%%%%žž

In addition to the measured data, the calculated FC force linewidths and FC force reduction are shown in Figs. \ref{Fig4}(a,b) by solid and dashed lines.
The calculations are performed for our experimental parameters using the theoretical models developed for the cw-induced radiation pressure force. 
By doing so, we considered a single comb mode participating in the interaction as a cw laser.
This consideration is reasonable given that the FC pulse repetition rate, $f_{rep}$, is much larger than the natural linewidth of the relevant transition, $\Gamma$, so the scattering rate of the neighboring comb modes is strongly reduced due to the ${(2{\pi}f_{rep}/\Gamma)^2}$ dependence \cite{santic2019, buhin2020, kielpinski2006}.

A detailed derivation of the average cw radiation pressure force resulting from the excitation of N atoms by a resonant laser can be found in \cite{bienaime2014}, and is given by:
\begin{equation}\label{force2}
	F = \frac{h k_0 \Gamma}{4\pi N} \int_{0}^{2\pi} \,d\phi \int_{0}^{\pi} \,d\theta \sin\theta (1-\cos\theta) I_s(\theta,\phi).
\end{equation}
$I_s(\theta,\phi)$ is the scattered far-field intensity, and $\theta, \phi$ determine the direction of the scattered photons.
Eq. (\ref{force2}) shows that the angular pattern of the scattered intensity uniquely determines the radiation pressure force, so to understand the force it is necessary to discuss different scattering mechanisms relevant to our experimental conditions.
Light scattering by an atomic cloud illuminated by a resonant laser can be decomposed into several contributions \cite{bachelard2016}: 
(a) The background radiation composed of diffuse scattering by all atoms. This contribution is incoherent in the sense that the phase of the scattered wave is random from one to another realization of atomic positions. 
(b) A forward lobe arising from the diffraction of the incident beam from the cloud, i.e. Mie scattering in the single scattering order. This contribution is coherent in the sense that the scattered wave has a well-defined phase.
(c) The coherent backscattering cone that arises from constructive interference during multiple scattering.

A full microscopic model built on a set of equations of $N$ coherently coupled dipoles (CD) can be used to calculate the scattering intensity \cite{bromley2016, zhu2016, jenkins2016}. This quantum model captures both incoherent and coherent contributions, including all scattering orders, and can be extended to include atomic motion. However, due to computational complexity, the CD model is limited to small samples, and as such is out of the scope of our paper. 

Another approach is to use semi-classical models to understand different scattering contributions and their influence on the radiation pressure force. 
To investigate these contributions, we will follow the analysis developed in \cite{bachelard2016} for the cw-induced force, and extend it beyond the limit of $b < 1$.

Diffuse scattering has two contributions to the force. 
The first one is called the shadow effect, and comes as a result of progressive attenuation of light intensity through the cloud due to diffuse scattering.
It can be explained by the Beer-Lambert law, i.e. the exponential decrease of the intensity results in broadening and reduction of the overall radiation pressure force.  
The force reduction arising from the shadow effect can be calculated from \cite{bachelard2016}:
\begin{equation}\label{force_shadow}
	\frac{F_{shadow}}{F_1}=\frac{\mathrm{Ein}(b)}{b},
\end{equation}
where $\mathrm{Ein}(b)$ is the entire function given by $\mathrm{Ein}(z)=\int_0^z dx(1-e^{-x})/x$, with $b=b_0/\left(1+4\delta^2/\Gamma^2\right)$, and $F_1$ the single-atom radiation pressure force.
We calculate $F_{shadow}$ spectra as a function of $b_0$ for our experimental parameters, from which we extract the $F_{shadow} (\delta)$ and $F_{shadow}$ linewidths. 
In order to obtain $F^1_{FC}(\delta)$, we fit Eq. (\ref{force_shadow}) to the measured FC force data as a function of $b_0$ for a given detuning $\delta$, $F^N_{FC}(\delta)$, with $F^1_{FC}(\delta)$ as a free fitting parameter. 
Thus determined $F^1_{FC}(\delta)$ is then used as a scaling factor for normalization of all $F^N_{FC}(\delta)$ forces shown in Figs. \ref{Fig4}(b).  
The calculated force linewidth and reduction as a function of $b_0$ resulting from the shadow effect are shown in Figs. \ref{Fig4}(a, b) by solid green lines.
The calculated values agree well with the measured data.

The second contribution to the force due to diffuse scattering is a consequence of the first one, i.e. since the light intensity is larger at the entrance of the cloud than at the exit, more light is scattered in the backward than in the forward direction. 
This causes an anisotropy of the emission pattern which slightly increases the radiation pressure force. 
This anisotropy can numerically be simulated using a random walk approach \cite{bachelard2016}, and becomes significant only at large optical thicknesses. 
Based on \cite{guerin2016josa}, we estimate that for the largest $b_0=20.8$ achieved in the experiment, the force including corrections due to anisotropy is around $10\%$ larger than the $F_{shadow}$, i.e. $F_{diffuse} = F_{shadow}+F_{anis} \approx 1.1 F_{shadow}$. 
This correction is within the uncertainty of the experimental data.

The contribution to the force due to diffraction of the incident beam can be calculated for clouds of small optical thickness \cite{rouabah2014} employing the Mie scattering approach \cite{bachelard2016}.
As stated in \cite{bender2010, bachelard2012}, this contribution is significant for very small atomic clouds ($kR\approx 10$) and for probe lasers tuned far off resonance, and it is therefore negligible for condition used in our experiment, i.e. large cloud ($kR > 1000$) and on-resonant excitation.

The coherent backscattering contribution can't be calculated using semi-classical models, but requires the full microscopic CD model \cite{bachelard2016, zhu2016, bienaime2011, javanainen1999}. 
However, as predicted in \cite{bachelard2016}, its contribution is also negligible for large clouds ($kR > 1000$) and on-resonant excitation such as in our experiment.

Because of its importance in the earlier experimental and theoretical papers \cite{bienaime2010, courteille2010, bienaime2011, bachelard2016}, we mention also an alternative approach used to investigate the radiation pressure force. It describes the force reduction as a consequence of coherent collective (i.e. cooperative) scattering of atomic dipoles. 
This cooperative contribution to the force can be calculated using a mean-field approach inspired by the timed-Dicke state (TDS). This model assumes that all atoms are driven by the unperturbed laser beam, i.e. the atoms acquire the phase of the laser and all have the same excitation probabilities.
It neglects reabsorption of photons by other atoms and works in conditions of small probe laser intensity or large detunings.
The TDS approach has become widely used in recent years, as it provides an explanation of experimental results on superradiance \cite{dicke1954, araujo2016}, a hallmark of cooperative effects.
Cooperative radiation pressure force, $F_{TDS}$, was studied in detail in \cite{bienaime2010, courteille2010}, and can be calculated from: 
\begin{equation}\label{force_cooperative}
	\frac{F_{TDS}}{F_1}= \frac{4\delta^2 + \Gamma^2}{4\delta^2 + (1+b_0/8)^2\Gamma^2}\left[1 + \frac{b_0}{16(k_0R)^2}\right],
\end{equation}
where $R$ is the cloud radius, and $F_1$ is the single atom force.
We calculate $F_{TDS}$ spectra as a function of $b_0$ for our experimental parameters, from which we extract the $F_{TDS} (\delta)$ and $F_{TDS}$ linewidths.
The calculated values are shown in Figs. \ref{Fig4}(a,b) by dashed violet lines. 
The TDS force agrees with measured data for small $b_0$ and coincides with the shadow effect curve up to $b_0\approx 3$ . However, at larger $b_0$ the TDS model predicts a linear increase of the force linewidth, which is not supported by our experimental results. 
TDS model predicts linear increase of the force linewidth with $b_0$, as it does not include multiple scattering effects that can induce the flattening of the force linewidth curve at large $b_0$ \cite{courteille2010}.
On the other hand, the good agreement of the force reduction calculated from the shadow and TDS models, Fig. \ref{Fig4}(b), even for intermediate $b_0$ explains why in earlier studies \cite{bienaime2010} the reduction of the force was ascribed to atomic cooperativity. 
However, the results of the force broadening given in Fig. \ref{Fig4}(a) clearly indicate that this agreement can be misleading, and point to the shadow effect as the dominant contribution to the force in dense atomic clouds. 
In the conditions when the atoms are resonantly excited by the frequency comb, the beam attenuation due to diffuse scattering is the dominant physical mechanism defining the radiation pressure force, and the atomic cooperativity effects are negligible.
This conclusion is in good agreement with measurements of superradiance, where superradiant enhancement was observed only for mid to large detunings, while tuning the probe close to resonance results in suppression of superradiant (cooperative) behavior \cite{araujo2016}. 

\section{Conclusion}
In conclusion, we have measured the frequency-comb-induced radiation pressure force acting on a cold $^{87}$Rb cloud as a function of the optical thickness of the cloud.  
We observed reduction and broadening of the frequency comb force as the optical thickness increases.
As the scattered intensity is directly mapped to the radiation pressure force, we discuss different scattering mechanisms and their contributions to the radiation pressure force.
For our experimental conditions, we show that a single scattering mechanism dominates the radiation pressure force.
It comes as a result of progressive attenuation of light intensity in the cloud due to diffuse scattering of light, i.e. the shadow effect. 
We also review the cooperative timed-Dicke state approach used in earlier experiments to investigate the radiation pressure force.
Theoretical models for cw radiation pressure force in the presence of shadow and cooperative effects developed in \cite{bachelard2016} and \cite{bienaime2010}, respectively, are used to describe the measured frequency comb force.
The measured and the calculated force broadening and reduction arising as results of the shadow effect are in good agreement.
The cooperative force agrees with measured data for small $b_0$, however the behaviour of force linewidth and force reduction at larger $b_0$ is not supported by the experiment.
This points to the shadow effect as the dominant contribution to the force modification in resonantly excited dense atomic clouds, i.e. a simple semiclassical model can be used to reproduce the measured force broadening and reduction. 
In order to observe the signature of the cooperative effects in the radiation pressure force, it would be necessary to work in the parametric regime where the beam attenuation due to diffuse scattering of light is negligible, such as large detunings from the atomic resonance.

Our results confirm the analogy between the cw and a single comb mode interaction, i.e. the influence of the off-resonance comb modes on the comb-atom interaction is minor and can be neglected, even in the case of increased optical thickness of the cloud.

The results presented in this paper contribute to the understanding of scattering of the frequency comb light by an ensemble of cold atoms, thus paving the way toward novel frequency comb applications in the field of cooling, quantum communication, and light-atom interfaces based on structured and disordered atomic systems.

\section{Acknowledgement}
The authors acknowledge support from the Croatian Science Foundation (Project Frequency comb cooling of atoms - IP-2018-01-9047).
The authors acknowledge Neven \v{S}anti\'{c} for reading the manuscript and providing constructive comments, as well as for his early contribution to the development of cold atoms experiment.
In addition, the authors acknowledge Ivor Kre\v{s}i\'{c} for the early contribution to the development of theoretical models as well as Grzegorz Kowzan and Piotr Mas\l{}owski for their contribution to the frequency comb stabilization.

%%%%%%%%%% If using BibTeX:
%\bibliography{bibl_3}

%

\end{document}